\documentclass[10pt, conference, compsocconf]{IEEEtran}
\ifCLASSINFOpdf
   \usepackage[pdftex]{graphicx}
   \graphicspath{{fig/}}
  \DeclareGraphicsExtensions{.pdf,.jpeg,.png}
\else
\fi
\usepackage{graphics}
\graphicspath{{fig/}}
\DeclareGraphicsExtensions{.pdf,.jpeg,.png,eps}

\usepackage{mdwmath}
\usepackage{mdwtab}
\usepackage{rotating}
\usepackage{stfloats}
\usepackage[font=small]{caption}
\usepackage[justification=centering]{caption}
\usepackage{multirow}
\usepackage{amssymb}
\usepackage[cmex10]{amsmath}
\usepackage[tight,footnotesize]{subfigure}
\usepackage[font=footnotesize]{subfig}
\usepackage{booktabs}
\usepackage{xcolor}
\usepackage{float}
\usepackage{boxedminipage}
\emergencystretch=\maxdimen
\usepackage[numbers,sort&compress]{natbib}

\begin{document}

%
% paper title
% can use linebreaks \\ within to get better formatting as desired
\title{Towards Cross-Project Defect Prediction with Imbalanced Feature Sets}

% author names and affiliations
% use a multiple column layout for up to two different
% affiliations
\author{\IEEEauthorblockN{Peng He\IEEEauthorrefmark{1}\IEEEauthorrefmark{2},
Bing Li\IEEEauthorrefmark{3}\IEEEauthorrefmark{4}, and
Yutao Ma\IEEEauthorrefmark{2}\IEEEauthorrefmark{4}}
\IEEEauthorblockA{\IEEEauthorrefmark{1}State Key Laboratory of Software Engineering, Wuhan University, Wuhan 430072, China}
\IEEEauthorblockA{\IEEEauthorrefmark{2}School of Computer, Wuhan University, Wuhan 430072, China}
\IEEEauthorblockA{\IEEEauthorrefmark{3}International School of Software, Wuhan University, Wuhan 430079, China}
\IEEEauthorblockA{\IEEEauthorrefmark{4}Research Center for Complex Network, Wuhan University, Wuhan 430072, China}
\IEEEauthorblockA{\{penghe, bingli, ytma\}@whu.edu.cn}

}
\maketitle

\begin{abstract}

 Cross-project defect prediction (CPDP) has been deemed as an emerging technology of software quality assurance, especially in new or inactive projects, and a few improved methods have been proposed to support better defect prediction. However, the regular CPDP always assumes that the features of training and test data are all identical. Hence, very little is known about whether the method for CPDP with imbalanced feature sets (CPDP-IFS) works well. Considering the diversity of defect data sets available on the Internet as well as the high cost of labeling data, to address the issue, in this paper we proposed a simple approach according to a distribution characteristic-based instance (object class) mapping, and demonstrated the validity of our method based on three public defect data sets (i.e., PROMISE, ReLink and AEEEM). Besides, the empirical results indicate that the hybrid model composed of CPDP and CPDP-IFS does improve the prediction performance of the regular CPDP to some extent.

\end{abstract}


%%
%% Start line numbering here if you want
%%
% \linenumbers

%% main text
\section{Introduction}
% no \IEEEPARstart
The importance of defect prediction has motivated numerous researchers to characterize various aspects of software quality by defining different prediction models. Most prior studies usually formulated such a problem as a supervised learning problem, that is, they trained defect predictors from the data of historical releases in the same project and predicted defects in the upcoming releases, or reported the results of cross-validation on the same data set \cite{He:An}, which is referred to as Within-Project Defect Prediction (WPDP). However, it is not always practical to collect sufficient historical data in new or inactive projects.

Nowadays, due to sufficient and freely available defect data from other projects, researchers in this field have been inspired to overcome the problem by applying the predictors built for one project to others \cite{Zimmermann:Cross,Briand:Assessing,Ma: Transfer}. This type of predictions is named as Cross-Project Defect Prediction (CPDP). The objective of CPDP is to predict defects in a project using the prediction model trained from the labelled defect data of other projects. Until now, the feasibility and potential usefulness of CPDP with a number of software metrics has been demonstrated \cite{Rahman:Recalling,He:An}.

\emph{\textbf{Motivation}}: Unfortunately, to the best of our knowledge, all existing CPDP models were built based on a rigorous hypothesis that training and test data must have the same set of software metrics (also known as features). Due to different data sources, many public defect data sets, such as the projects in ReLink\footnote{http://www.cse.ust.hk/$\sim$scc/ReLink.htm},  AEEEM\footnote{http://bug.inf.usi.ch/} and PROMISE\footnote{http://promisedata.org/}, consist of different software metrics. Moreover, different data contributors may provide various sets of metrics for the same project. If we want to predict software defects of a project in ReLink, the existing CPDP prediction methods seem to be useless when only the labelled defect data from AEEEM is available at hand. Because of imbalanced feature sets between the source and target projects, we have to re-collect data using the same set of metrics as that of the target project. Therefore, there is no doubt that the time-consuming data collection, annotation and validation is redundant and trivial if the CPDP with imbalanced feature sets (CPDP-IFS) can be realized.

So far, prior studies on CPDP have investigated how to select the appropriate training data for CPDP \cite{Peters:Better,Turhan:On} and how to reduce the dimensions of feature set by feature selection techniques \cite{Wang: A2,Liu:Toward,Peng: Feature}.  However, as far as we know, there are no relevant studies to discuss the issue. That is, the feasibility of CPDP with different sets of metrics for training and test data is still an open challenge. Thus, can CPDP-IFS achieve a comparable (or even better) result compared with the regular CPDP? If so, on the one hand, it will improve the utilization of available defect data and reduce the effort of data acquisition, annotation and validation; on the other hand, it can enhance the generality of the regular CPDP.

\emph{\textbf{Idea}}: Unlike the regular CPDP, the essential characteristic of CPDP-IFS in this paper is independent of the number and type of metrics for training and test data. Assuming that an instance (object class) can be regarded as a vector of metrics, the vectors with different lengths may have the same or similar statistical distribution of numerical values of metrics. Additionally, most of the instances whose metrics are all within the normal range rarely contain bugs. Instead, an instance is more likely to be defective in case of abnormal distribution characteristics (such as mean and variance) caused by one or more particularly prominent metrics. Hence, the distribution characteristics of metrics may be a potential indictor for software defect-proneness.

In this paper, we proposed a new approach to CPDP-IFS, which is based on the assumption that an instance (probably) tends to contain bugs if its distribution characteristics of metrics are similar to those of defective instances. In short, we projected the instances from both source and target projects onto a latent space composed of distribution indicators of their metrics, and applied the regular CPDP to the converted data with the same features. Our contributions to the current state of research are summarized as follows:
\begin{itemize}
\item We formulated and presented a simple distribution characteristic-based instance mapping approach to CPDP-IFS, which aims to address the imbalance of metric sets in CPDP.
\item Based on three public data sets, we first validated the feasibility of our method for CPDP-IFS using statistical analysis methods.
\item We further built a hybrid model that consists of CPDP and CPDP-IFS, and found that it was able to significantly improve the performance of defect prediction in some specific scenarios.
\end{itemize}

The rest of this paper is organized as follows. Section \uppercase\expandafter{\romannumeral2} is a review of related literature. Sections \uppercase\expandafter{\romannumeral3} describes the problem we attempted to address and our approach. Sections \uppercase\expandafter{\romannumeral4} and \uppercase\expandafter{\romannumeral5} show the detailed experimental setups and analyze the primary results, respectively. Some threats to validity that could affect our study are presented in Section \uppercase\expandafter{\romannumeral6}. Finally, Section \uppercase\expandafter{\romannumeral7} concludes the paper and presents the agenda for future work.

\section{Related Work}
\subsection{Cross-Project Defect Prediction}
To the best of our knowledge, prior studies focused mainly on validating the feasibility of CPDP. For example, Briand \emph{et al.} \cite{Briand:Assessing} first applied the model trained from the Xpose project to predicting the Jwriter project, and validated that such a CPDP model did perform better than the random model. In \cite{Zimmermann:Cross,Rahman:Recalling}, the authors investigated the performance of CPDP in terms of a large scale experiment on data vs. domain vs. process and cost-sensitive analysis, respectively. Furthermore, He \emph{et al.} \cite{He:An} validated the feasibility of CPDP based on a practical performance criterion (precision greater than 0.5 and recall greater than 0.7), and they also proposed an approach to automatically selecting suitable training data for those projects without local data.

Considering the choice of training data from other projects, Turhan \emph{et al.} \cite{Turhan:On} proposed a nearest-neighbor filtering technique to filter out the irrelevancies in cross-project data, and they also found that only $10\%$ of the historical data could make mixed project predictions perform as well as WPDP models \cite{Turhan:Empirical}. An improved instance-level filtering strategy was then proposed in \cite{Peters:Better}; on the other hand, Herbold \cite{Herbold:Training} proposed two methods for selecting the proper training data at the level of release. The results demonstrated that their selection methods improved the achieved success rate significantly, although the quality of the results was still unable to compete with that of WPDP.

%However, relatively little attention has been paid to empirically exploring the feasibility of a predictor where the metric set of its source project is different from that of target project. Moreover, very little is known about whether the predictors built with different metric sets are able to enhance the original CPDP.

\subsection{Transfer Learning Techniques}
In machine learning, transfer learning techniques have attracted great attention over the last several years \citep{Pan: A}, and the successful applications include effort estimation \citep{Kocagune: Transfer}, text classification \citep{Xue: Topic}, name-entity recognition \citep{Arnold: A}, natural language processing \citep{Pan: Cross}, etc. Recently, it has been proven to be appropriate for CPDP \citep{Nam:Transfer}, since the problem setting of CPDP is related to the adaptation setting that a classifier in the target project is built using the training data from those relevant source projects. The typical applications of defect prediction include Transfer Na\"{\i}ve Bayes (TNB) \cite{Ma: Transfer} and TCA (Transfer Component Analysis) \cite{Nam:Transfer}. In this paper, we conducted two types of experiments on CPDP (without transfer learning and with TCA) to investigate the feasibility and generality of our approach.

\subsection{Software Metrics}

Shin \emph{et al.} \cite{Shin: Evaluating} investigated whether source code and development histories were discriminative and predictive of vulnerable code locations. Marco \emph{et al.} \cite{Marco:Evaluating} conducted three experiments with process metrics, previous defects, source code metrics, entropy of changes, churn, etc., to evaluate different defect prediction approaches. In \cite{Zimmermann:Predicting,Tosun:Validation,Premraj:Network}, the authors leveraged social network metrics derived from dependency relationships to predict defects. More studies can be found in literature \cite{Arisholma:A,Nagappan:Using,Menzies:Data,Yu:Experience}. Actually, different software metrics measure various aspects of software. Due to the difference of metric sets, most of defect data sets provided in prior studies cannot be directly used to validate other work, and they are even unsuitable for the regular CPDP. However, in fact, these labelled defect data sets may be very valuable for CPDP if we can find an appropriate approach to preprocessing and transforming them.

\section{Problem and Approach}
\begin{figure}
\centering
\includegraphics[width=3in,height=1.8in]{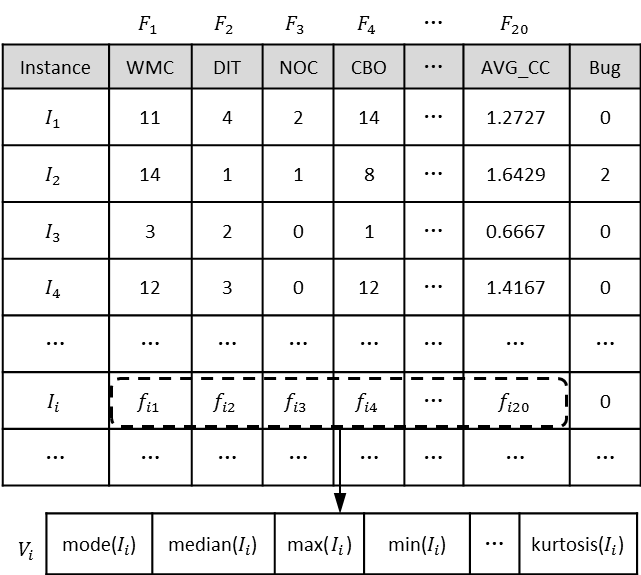}
\caption{An example of the Ant project's defect data set: instances ($I$), distribution characteristics ($V$) and features ($F$).}
\label{release}
\end{figure}

CPDP is defined as follows: Given a source project $P_{S}$ and a target project $P_{T}$, CPDP aims to achieve the target prediction in $P_{T}$ using the knowledge extracted from $P_{S}$, where $P_{T}\neq P_{S}$. Assuming that the source and target projects have the same set of features, they may differ in feature distribution characteristics. The goal of CPDP is to learn a model from the selected source projects (training data) and apply the learned model to the target project (test data). In our context, a project $P$, which contains $m$ instances, is represented as $P=\{I_{1},I_{2},\cdots,I_{m}\}$. An instance can be represented as $I_{i}=\{f_{i1}, f_{i2},\cdots,f_{in}\}$, where $f_{ij}$ is the $j^{th}$ feature value of the instance $I_{i}$, and $n$ is the number of features. A distribution characteristic vector of the instance $I_{i}$ can be formulated as $V_i=\{c_{i1},c_{i2},\cdots,c_{ik}\}$, where $k$ is the number of distribution characteristics, e.g., $mean$, $median$ and $variance$ (see Figure \ref{release}).

\subsection{Problem Analysis of the Regular CPDP }
For a newly created or inactive project, one of the easiest methods of defect prediction is CPDP, that is, one can directly train a prediction model with the defect data from other existing projects. Unfortunately, due to different provenances of these existing public data sets, they usually consist of different sets of metrics, and the scales of these metric sets are also varied. As a consequence, this increases the burden of data acquisition and validation of metrics because of the basic hypothesis that the target and source projects have the same set of features in the regular CPDP.

Once common features exist between the source and target projects, a simplest method to deal with the issue is to use the intersection between feature sets of the training and test data. If there is no intersection, a reasonable method is to perform a transformation process, so as to ensure that the feature sets of the source and target projects are still identical. To the best of our knowledge, the transfer learning technique, a state-of-the-art feature extraction technique, has been applied to CPDP frequently by some researchers. The motivation behind transfer learning is that some common latent factors may exist between the source and target projects, even though the observed features are different. Through mapping the source and target projects onto a latent space, the difference between them can be reduced and the original data structures can be preserved. As a result, the latent space spanned by these latent factors can be used as a bridge for CPDP.

\begin{figure}
\centering
\includegraphics[width=3in,height=1.8in]{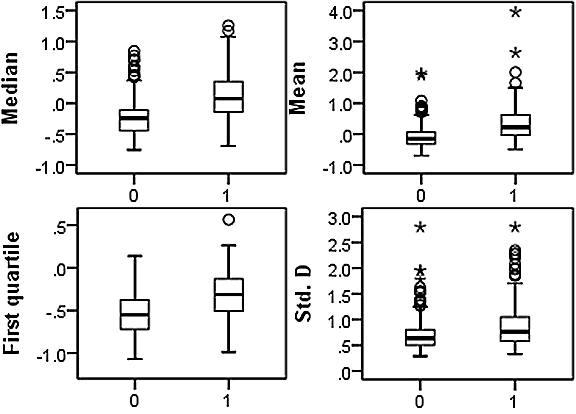}
\caption{The standardized boxplots of four indictors of feature values between defective (1) and defect-free (0) instances.}
\label{indicator}
\end{figure}

Inspired by the idea of transfer learning, we conducted a small-scale experiment on the Ant project to test the feasibility of distribution characteristic-based instance mapping for CPDP-IFS. For each instance $I$ of this project (see Table \ref{projects}), we calculated its distribution characteristic vector $V$ in terms of $Mean$, $Median$, $First\quad Quartile$ and $Standard\quad Deviation$. Interestingly, the result shows that defective instances tend to have higher $Mean$, $Median$ and $First\quad Quartile$ values than those defect-free ones, and that the fluctuation of their feature values is also greater according to $Standard\quad Deviation$ (see Figure \ref{indicator}). The observation implies that distribution characteristics seem to be proper components of the latent space we want. Therefore, our feasible solution is to project the instances of the source and target projects onto a common latent space which is related to the distribution characteristics of feature values. We then apply the regular CPDP to the converted data in the common space.

\emph{\textbf{Importance:}} In this paper, CPDP represents the regular cross-project defect prediction, where the source and target projects possess the same set of metrics. CPDP-IFS is actually a specific type of CPDP, where the source and target projects have different metric sets. Of course, the study on CPDP-IFS can improve the generality and practicality of the regular CPDP, which is the main motivation of this paper.

\subsection{Research Questions}
According to the problem analysis, we attempt to find empirical evidence that addresses the following three research questions in this paper:

\begin{itemize}

  \item \emph{RQ1: Does our method for CPDP-IFS perform better than the intersection-based method?}\\
      As mentioned before, there are two simple approaches to CPDP-IFS. So, we need to carry out a comparison of the homogeneous methods for building a common latent space of identical features between the source and target projects.
  \item \emph{RQ2: Is the performance of our method for CPDP-IFS comparable to that of CPDP?}\\
      To validate the feasibility of CPDP-IFS, we also need to perform a vertical comparison between it and CPDP. If the results of our approach to CPDP-IFS are significantly worse than those of the regular CPDP, the feasibility of the approach should be questioned.
  \item \emph{RQ3: Can the hybrid model composed of CPDP-IFS and CPDP improve the performance of CPDP?}\\
    If CPDP-IFS is feasible and can be an important supplement to the regular CPDP, we are eager to know whether a blend combining them can achieve better performance.

\end{itemize}

\subsection{Our Approach to CPDP-IFS}\label{trans}

As shown in Figure \ref{release}, since different features have different scales in a project, the feature values have to be pre-processed to avoid comparing the largest ``ant" with the smallest ``elephant". In addition, some prior studies have suggested that a predictor's performance might be improved by applying a proper filter to numerical values when the distribution of values for a feature is highly skewed \cite{Turhan:Empirical}. Therefore, in this paper we accomplish the procedure of CPDP-IFS with the following three steps:

\begin{enumerate}
  \item[(1)] \emph{Preprocessing}: Applying a preprocessing method such as logarithmic filter to numerical values if necessary, and normalizing each feature $F_i$ by the $z-score$ method. Note that the logarithmic filter is optional and other normalization methods also can be used.
  \item[(2)] \emph{Mapping}: Projecting the instances of source and target projects onto a latent space according to the given indicators, so a project $P=\{I_{1},I_{2},\cdots,I_{m}\}$ will be transformed as $P^{'}=\{V_{1},V_{2},\cdots,V_{m}\}$ in our context. Due to the limit of space, only 5 out of 16 typical indicators used to represent distribution characteristics are listed in Table \ref{characteristic}.
  \item[(3)] \emph{Learning}: After the mapping, one can perform the regular CPDP for the converted data of the projects from different data sets.
\end{enumerate}

\begin{table}\footnotesize
\centering
\caption{Descriptions of 5 indicators. (For a detailed description of all indicators we used, please refer to He \emph{et al}. \cite{He:An})}\label{characteristic}
\begin{tabular}[c]{p{20mm}p{60mm}}
  \hline
  % after \\: \hline or \cline{col1-col2} \cline{col3-col4} ...
  Indicator & Description \\  \hline
  Median  & The numerical value separating the higher half of a population from the lower half   \\
  Mean  & The average value of samples in a population \\
  Min  & The least value in a population \\
  Max  & The greatest value in a population \\
  Variance & The arithmetic mean of the squared deviation of the Mean to values of cases \\
  %Standard Deviation  & The square root of the Variance \\
%  Mode & The value that occurs most frequently in a population \\
%  Harmonic Mean & The reciprocal of the arithmetic mean of the reciprocals \\
%  Range & The deviation of the Max to the Min \\
%  Variation Ratio & The proportion of cases that are not the mode \\
%  First Quartile & The value cutting off $25\%$ lowest cases in a population\\
%  Third Quartile & The value cutting off $75\%$ lowest cases in a population\\
%  Interquartile Range & The deviation of the First Quartile to the Third Quartile\\
%  Coefficient of Variation & The ratio of the Standard Deviation to the Mean \\
%  Skewness & A measure of the asymmetry of a population \\
%  Kurtosis & A measure of the peakedness of a population \\
  \hline
\end{tabular}
\end{table}

\section{Experimental Setup}
\subsection{Data Collection}
 For our experiments, we used three on-line public defect data sets (i.e., PROMISE \cite{Jureczko:Using}, ReLink \cite{Wu:ReLink}, and AEEEM \cite{D¡¯Ambros: An}), including a total of 11 different projects. Detailed information of the three data sets is summarized in Table \ref{projects}, where $\#$ \emph{instances}, $\#$ \emph{defects} and $\#$ \emph{metrics} are the numbers of instances, defects and metrics, respectively. Each instance in these public data sets represents a class file and consists of two parts: independent variables related to software metrics and a dependent variable about defect. The number of instances varies from 56 to 1862, the defect ratio ranges from $2.9\%$ to $46.9\%$, and the size of metric sets is not less than 20.

\begin{table*}\footnotesize
\centering
\caption{Projects in the three data sets used in our experiments.}\label{projects}
\begin{tabular}{c|c|c|c|c|c}
  \hline
  % after \\: \hline or \cline{col1-col2} \cline{col3-col4} ...
 Data set&  Project  & Version & \# instances(files) & \# defects(\%) &\# metrics \\ \hline
  \multirow{3}*{PROMISE}  & Ant    & 1.7    &  745   &  166(22.3)  & 20 \\
         &  Camel   & 1.6    &  965   &  188 (19.5) & 20 \\
         & Xalan   & 2.6    &  885   &  411 (46.4) & 20 \\
  \hline
  \multirow{3}*{ReLink} & Apache HTTP Server (Apache)   & 2.0    &  194   &  91(46.9)  & 40 \\
  & OpenIntents Safe (Safe)  & R1088-2073    &  56   &  16 (28.6) & 40 \\
  & ZXing   & 1.6    &  399   &  83 (20.8) & 40 \\
  \hline
  \multirow{5}*{AEEEM} & Equinox     & 1.1.2005-6.25.2008 &  324   &  129(39.8)  & 76 \\
  & Eclipse JDT core(Eclipse) & 1.1.2005-6.17.2008   &  997   &  206 (20.7) & 76 \\
  & Apache Lucence (Lucence)  & 1.1.2005-10.8.2008 &  692   &  20 (2.9) & 76 \\
  & Mylyn  & 1.17.2005-3.17.2009 &  1862   &  245 (13.2) & 76 \\
  & Eclipse PDE UI (Pde)  & 1.1.2005-9.11.2008   &  1497   &  209 (14.0) & 76 \\
  \hline
\end{tabular}
\end{table*}

\begin{figure*}
\centering
\includegraphics[width=7in,height=2in]{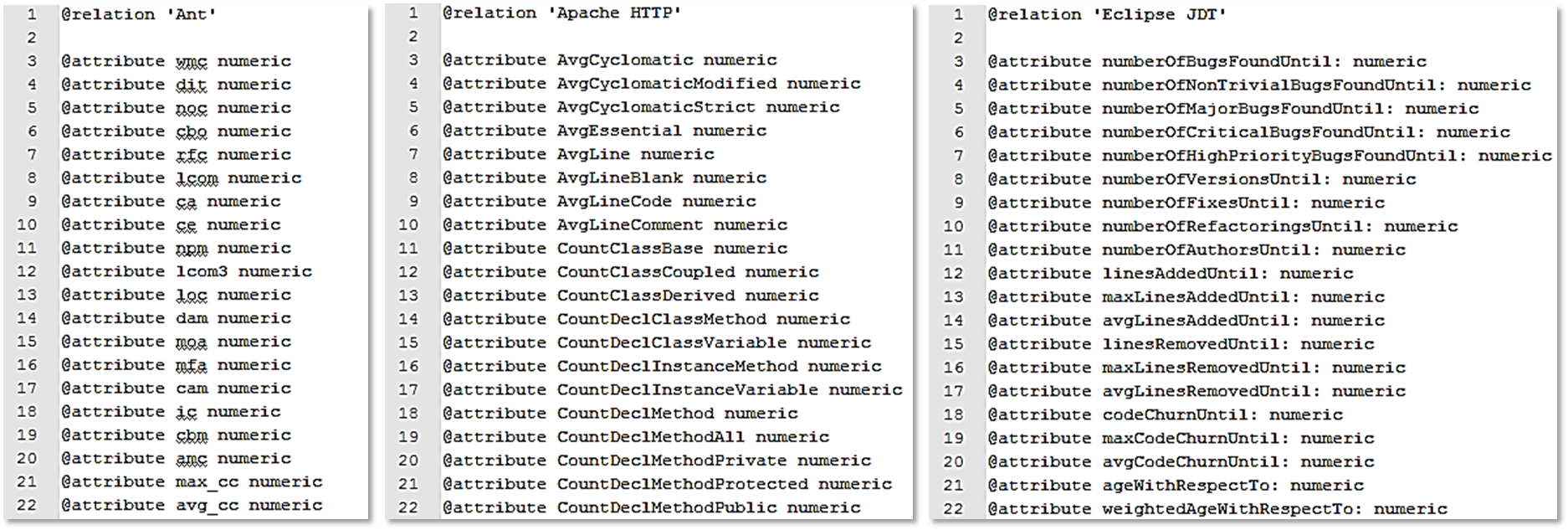}
\caption{A snapshot of the metrics used in the three data sets. (The size and type of metric sets: $Size_{PROMISE}<Size_{ReLink}$ $<Size_{AEEEM}$ and $Type_{PROMISE} \cap Type_{ReLink}={\o}$, $Type_{PROMISE} \cap Type_{AEEEM}\neq {\o}$, $Type_{AEEEM} \cap Type_{ReLink}={\o}$).}
\label{metrics}
\end{figure*}

The first data set, PROMISE, was collected by Jureczko and Spinellis \cite{Jureczko:Using}. The information of defects in PROMISE has been validated and used in several prior studies. The second data set, ReLink, was collected by Wu \emph{et al}. \cite{Wu:ReLink} and has been manually verified and corrected by the authors. Note that, only the common metrics of the three projects in ReLink, a total of 40 metrics, were used in this paper. The third data set, AEEEM, was collected by D' Ambros \emph{et al}. \cite{D¡¯Ambros: An}. This data set consists of 76 metrics: 17 source code metrics, 15 change metrics, 5 previous defect metrics, 5 entropy-of-change metrics, 17 entropy-of-source-code metrics, and 17 churn-of-source-code metrics. For the needs of our experiments, the sizes and types of metric sets are varied among the three data sets. Figure \ref{metrics} shows a snapshot of the metric sets.

\subsection{Experimental Design}
\begin{figure}
\centering
\includegraphics[width=3.2in,height=1.2in]{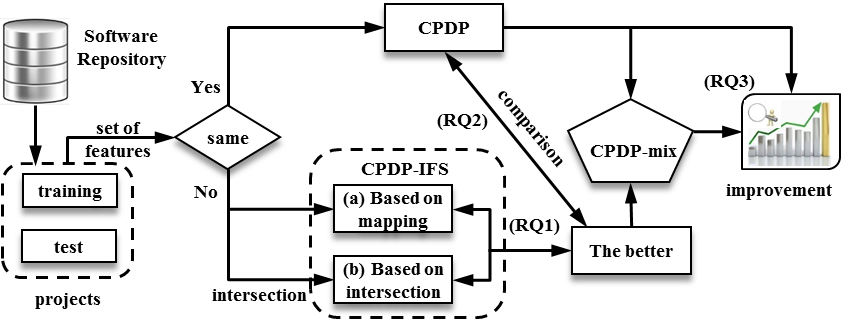}
\caption{The framework of our experiments}
\label{Fig.framework}
\end{figure}

In this subsection, we present the experimental design in detail, including three types of cross-project defect predictions. Figure \ref{Fig.framework} shows the entire framework of our experiments. First, if training and test data have different feature sets, we have two choices to realize CPDP-IFS: our method based on distribution characteristic and the method based on intersection; otherwise, we choose the regular CPDP. Second, we compare the two methods for CPDP-IFS, and use the better one as the recommended approach to CPDP-IFS to validate its feasibility by comparing with the regular CPDP. Third, to improve prediction accuracy, we further integrate CPDP-IFS into the regular CPDP to predict defects regardless of the original assumption of CPDP. Besides, we also attempt to provide some practical guidelines for determining appropriate source projects during performing CPDP-IFS.

\emph{1) Two settings for CPDP:} In our context, predictors are built in two settings: CPDP without transfer learning and CPDP with transfer learning (i.e., TCA \cite{Pan: Domain}). We use $CPDP_{pure}$ and $CPDP_{tca}$ to label the two types of defect prediction models respectively. Before building a predictor, we have to set up the source and target projects. For example, PROMISE has 6 combinations: Ant$\Leftrightarrow$ Camel, Ant$\Leftrightarrow$ Xalan, Camel$\Leftrightarrow$ Xalan. We need to build a predictor with the project at one side of the arrow and apply the predictor to the project at the other side. In the same manner, we also identify all 6 and 20 combinations in ReLink and AEEEM, respectively.

\emph{2) Two methods for CPDP-IFS:} The main task of this paper is to investigate the feasibility of the cross-project predictions between the data sets with different metric sets. In this paper, we introduce two simple methods for CPDP-IFS to address the issue. One is based on distribution characteristics (labeled as $CPDP-IFS^{our}$), the other is based on intersection (labeled as $CPDP-IFS^{min}$). Then, we train a predictor with the converted data from source projects in the two settings separately, and use it to predict defects in the transformed target project. During this process, the regular CPDP predictions must be excluded when the source and target projects are from the same data set. For instance, for Xalan, the predictions Ant$\rightarrow$ Xalan and Camel$\rightarrow$ Xalan cannot be included in this experiment.

\emph{3) The hybrid model (CPDP-mix):} To our knowledge, although the feasibility of CPDP has been demonstrated, the overall performance is still not good enough in practice \cite{Rahman:Recalling}. With the help of CPDP-IFS, we further analyze the generality and practicality of CPDP to investigate whether \emph{CPDP-mix} can improve the prediction performance of the regular CPDP. Thereby, we re-compare \emph{CPDP-mix} with the original CPDP in terms of prediction performance.

\subsection{Classifier and Evaluation Measures}
As one of the commonly-used classifiers for cross-project defect predictions, logistic regression has been used in several prior studies \cite{He:An,Nam:Transfer,Pan: Domain}. Specifically, in this paper we used the algorithm of logistic regression implemented in Weka\footnote{http://www.cs.waikato.ac.nz/ml/weka/} and the default parameter settings specified in Weka.

In general, there are trade-offs between \emph{precision} and \emph{recall}, so we adopt \emph{f-measure} to evaluate prediction performance as other researchers did in prior studies \cite{Nam:Transfer}. As we know, a binary classification prediction will produce four possible results: \emph{false positive (FP)}, \emph{false negative (FN)}, \emph{true positive (TP)} and \emph{true negative (TN)}. The followings are used to describe the \emph{precision}, \emph{recall}, and \emph{f-measure}:

%\begin{table}\small
%\centering
%\caption{The measures used in our experiments}\label{measures}
%\begin{tabular}{c|c}
%  \hline
%  % after \\: \hline or \cline{col1-col2} \cline{col3-col4} ...
%  \textbf{Measures} & \textbf{Classifiers} \\ \hline
%  \multicolumn{1}{c|}{$Recall(pd)$ = $\frac{TP}{TP+FN}$ } & \multicolumn{1}{c}{J48 Decision Tree (J48)}\\
%  \multicolumn{1}{c|}{$Precision(prec)$  =  $\frac{TP}{TP+FP}$} & \multicolumn{1}{c}{Logistic Regression (LR)} \\
%  \multicolumn{1}{c|}{$pf=\frac{FP}{FP+TN}$ } & \multicolumn{1}{c}{Na\"{\i}ve Bayes (NB)} \\
%  \multicolumn{1}{c|}{$f$-measure =$\frac{2*pd*prec}{pd+prec}$ }& \multicolumn{1}{c}{Support Vector Machine (SVM)} \\
%  \multicolumn{1}{c|}{$g$-measure =$ \frac{2*pd(1-pf)}{pd+(1-pf)}$} & \multicolumn{1}{c}{Random Forest (RF)} \\
%  \hline
%\end{tabular}
%\end{table}

\begin{itemize}
  \item \emph{precision} addresses how many of the defect-prone instances returned by a model are actually defective. The best precision value is 1. The higher the precision is, the fewer false positives (i.e., defect-free elements incorrectly classified as defective ones) exist:
    \begin{equation}\label{Eq.precision}\small
        precision=\frac{TP}{TP+FP}.
    \end{equation}
  \item \emph{recall} addresses how many of the defect-prone instances are actually returned by a model. The best recall value is 1. The higher the recall is, the lower the number of false negatives (i.e., defective elements missed by the model) is:
     \begin{equation}\label{Eq.recall}\small
        recall=\frac{TP}{TP+FN}.
     \end{equation}
  \item \emph{f-measure} considers both \emph{precision} and \emph{recall} to compute the accuracy, which can be interpreted as a weighted average of \emph{precision} and \emph{recall}. The value of \emph{f-measure} ranges from 0 to 1, with values closer to 1 indicating better performance for classification results.
     \begin{equation}\label{Eq.F}\small
       f-measure=\frac{2*precision*recall}{precision+recall}.
     \end{equation}
\end{itemize}

%The area under the ROC curve (AUC) is the portion of the area of a unit square that is equal to the probability that a classifier will identify a randomly chosen defective class higher than a randomly chosen defect-free one \cite{Fawcett: An}. An AUC less than 0.5 means a very low true positive rate and high false alarm rate. Therefore, we also used AUC to evaluate the performance of the predictors built with simplified metric set in the following experiments.

\section{Experimental Results}

%In this section, we report the primary results so as to answer the three research questions formulated in Section 3.2.

\subsection{Does our method for CPDP-IFS perform better than the intersection-based method?}

\begin{figure*}
\centering
\includegraphics[width=5in,height=1.5in]{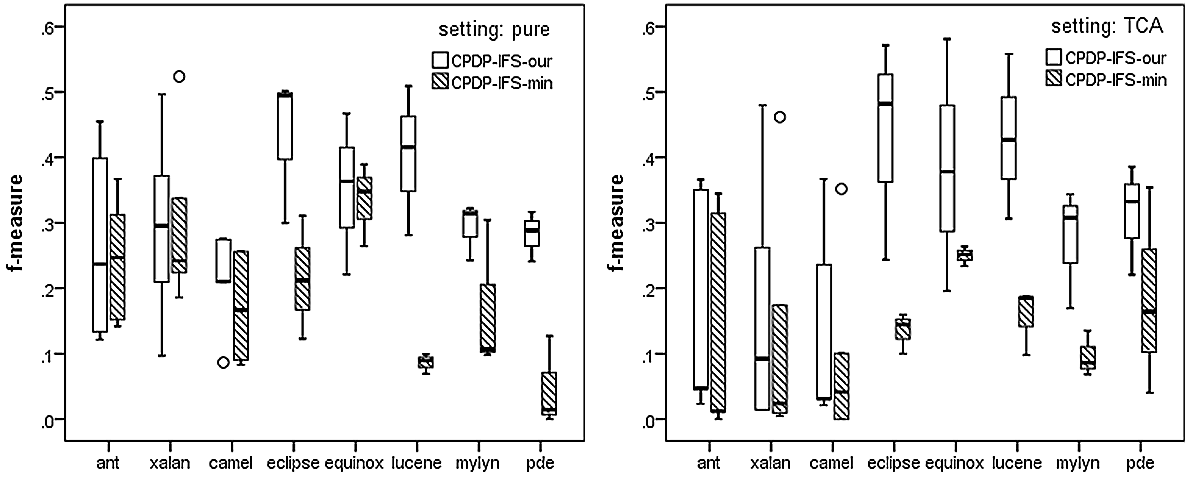}
\caption{The standardized boxplots of \emph{f-measure} values obtained by the two methods in two given settings. From the bottom to the top of a standardized box plot: minimum, first quartile, median, third quartile and maximum. The outliers are plotted as circles.}
\label{ab}
\end{figure*}

First of all, we compared the prediction performance of the two CPDP-IFS methods in two given settings. In Figure \ref{ab}, it is clear that the median values of our approach are, in general, larger than those of $CPDP-IFS^{min}$ in both settings in terms of \emph{f-measure}, though there are two exceptions: Ant (setting: pure) and Camel (setting: tca). In particular, for the data set AEEEM, the improvement in performance is more significant. Note that, there are no common metrics between ReLink and the other two data sets so that we only analyzed the 8 projects included in PROMISE and AEEEM.

\begin{table}[H]\footnotesize
\newcommand{\tabincell}[2]{\begin{tabular}{@{}#1@{}}#2\end{tabular}}
\centering
\caption{A performance comparison between the two methods in the $CPDP-IFS_{pure}$ setting according to the Wilcoxon signed-rank test ($p-value=0.05$) and Cliff's Delta ($d$). }\label{CPDP}
\begin{tabular}{c|c|c|c|c|c} \hline
\multirow{2}*{Target} &\multicolumn{2}{c|}{$CPDP-IFS^{our}_{pure}$} &\multicolumn{2}{c|}{$CPDP-IFS^{min}_{pure}$} &\multirow{2}*{\tabincell{c}{$Sig.p$ \\($d$)}} \\
 \cline{2-5}
  & Source & Value & Source & Value &
   \\ \hline
    Ant & Eclipse & 0.45 & Equinox & 0.37 & \multirow{8}*{\tabincell{c}{0.036\\ (0.5)}}\\
    Xalan   & Equinox & 0.50 & Equinox & 0.52 &  \\
    Camel   & Equinox & 0.28 & Equinox & 0.26 & \\
    Eclipse   & Ant & 0.50 & Ant & 0.31 &  \\
    Equinox  & Xalan & 0.47 & Xalan & 0.39 &  \\
    Lucene  & Camel & 0.51 & Camel & 0.10 & \\
    Mylyn   & Ant & 0.32 & Ant & 0.30 & \\
    Pde & Ant & 0.32 & Ant & 0.13 &\\ \hline
\end{tabular}
\end{table}

Then, we compared the best prediction results of the two methods according to statistical analysis methods. Table \ref{CPDP} lists their corresponding source projects and the maximum \emph{f-measure} values achieved in the $CPDP-IFS_{pure}$ setting among 30 ($3*5+5*3=30$) predictions. For the first project Ant, its optimal source projects are different according to the two methods. Meanwhile, the $p$-value indicates that we have to reject the \emph{null hypothesis} that the two sets of values are drawn from the same distribution in terms of the Wilcoxon signed-rank test ($p$-value$=0.036<0.05$). That is, there is a statistically significant difference between $CPDP-IFS^{our}_{pure}$ and $CPDP-IFS^{min}_{pure}$ when only considering the best results. As a non-parametric effect size measure that quantifies the amount of difference between two groups of observations beyond $p$-values interpretation \cite{Macbeth: Cliff}, the positive Cliff's Delta ($d=0.5$) means that the left-hand values are higher than the right-hand ones in our context, i.e., the effect size of our approach is larger than that of $CPDP-IFS^{min}_{pure}$. This suggests that our approach is more useful for the $CPDP-IFS$ without transfer learning. For example, for the Lucene project, the best performance was increased by 0.41 using our method.

\begin{table}[H]\footnotesize
\newcommand{\tabincell}[2]{\begin{tabular}{@{}#1@{}}#2\end{tabular}}
\centering
\caption{A performance comparison between the two methods in the $CPDP-IFS_{tca}$ setting according to the Wilcoxon signed-rank test ($p-value=0.05$) and Cliff's Delta ($d$). }\label{TCA}
\begin{tabular}{c|c|c|c|c|c} \hline
\multirow{2}*{Target} &\multicolumn{2}{c|}{$CPDP-IFS^{our}_{tca}$} &\multicolumn{2}{c|}{$CPDP-IFS^{min}_{tca}$} &\multirow{2}*{\tabincell{c}{$Sig.p$ \\($d$)}} \\
 \cline{2-5}
  & Source & Value & Source & Value &
   \\ \hline
    Ant & Equinox & 0.37 & Eclipse & 0.35 & \multirow{8}*{\tabincell{c}{0.012\\ (0.781)}}\\
    Xalan   & Equinox & 0.48 & Equinox & 0.46 &  \\
    Camel   & Equinox & 0.37 & Equinox & 0.35 & \\
    Eclipse   & Xalan & 0.57 & Camel & 0.16 &  \\
    Equinox  & Xalan & 0.58 & Xalan & 0.26 &  \\
    Lucene  & Camel & 0.56 & Camel & 0.19 & \\
    Mylyn   & Xalan & 0.34 & Ant & 0.14 & \\
    Pde & Xalan & 0.39 & Ant & 0.35 &\\ \hline
\end{tabular}
\end{table}

In the $CPDP-IFS_{tca}$ setting, Table \ref{TCA} shows very similar results. Besides Ant, Eclipse also has different optimal source projects when using the two methods. There is a statistically significant difference between $CPDP-IFS^{our}_{tca}$ and $CPDP-IFS^{min}_{tca}$, indicated by $p$-value$=0.012<0.05$. According to the Cliff's Delta ($d=0.781$), the effect size of $CPDP-IFS^{our}_{tca}$ is also larger than that of $CPDP-IFS^{min}_{tca}$, and the disparity becomes quite remarkable. For the Eclipse project, the best performance was also increased by 0.41 using our method.

Despite the simplicity and usability of the intersection-based method for CPDP-IFS, it becomes useless when there are no common metrics between two projects. Remarkably, our approach not only performs better than $CPDP-IFS^{min}$, but also is more general for CPDP-IFS. Therefore, our approach should be preferentially recommended to solve the problem of imbalanced feature sets. In other words, the distribution characteristics of normalized feature values of instances are more suitable to preserve the actual defect information than the intersection of common features. For defective instances, a possible explanation is that the values of some commonly-used metrics may usually change (become larger or smaller) due to the maintenance and evolution process, in particular when one or more defects are repaired by different developers. According to the finding, we will test the feasibility of our approach to CPDP-IFS compared with the regular CPDP in the following experiment.

\subsection{Is the performance of our method for CPDP-IFS comparable to that of CPDP?}

In this experiment, $CPDP_{pure}$ and $CPDP_{tca}$ were used as two baselines for the regular CPDP predictions. For the three data sets, we conducted 32 ($6+6+20=32$) CPDP predictions and 78 ($3*8+3*8+5*6=78$) CPDP-IFS predictions, and selected 11 best results among these predictions to compare the performance of CPDP and CPDP-IFS. Thus, based on the \emph{null hypothesis} that there is no significant difference between CPDP-IFS and CPDP (i.e., $H_0: \mu_{CPDP-IFS}-\mu_{CPDP}=0$), we made a comparison between them in terms of the Wilcoxon signed-rank test and Cliff's effect size (see Table \ref{CFDP-CPDP}). The $p$-values yielded by the test suggest that the performance of CPDP-IFS is comparable to that of the regular CPDP. For example, the $p$-value $0.906$ between $CPDP-IFS_{pure}$ and $CPDP_{pure}$ indicates that their best prediction results are very similar in terms of \emph{f-measure}. Additionally, non-negative Cliff's Delta $d$ values show the superiority of CPDP-IFS over CPDP, suggesting the feasibility of our method. Note that, a positive $d$ implies that the effect size of CPDP-IFS is greater than that of CPDP.

\begin{figure}
\centering
\includegraphics[width=3in,height=1.8in]{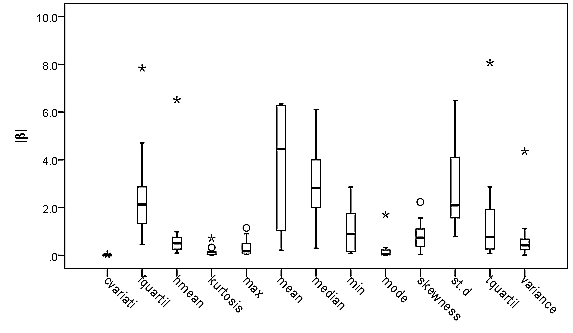}
\caption{The distribution of the absolute coefficient $|\beta|$ for each indicator in the 11 projects with Logistical Regression Analysis. }
\label{contri}
\end{figure}

For the predictions in the $CPDP-IFS_{pure}$ setting without feature selection, we also performed a Logistic Regression analysis on each transformed target project to distinguish the contribution of each distribution indicator to a predictor's performance. Figure \ref{contri} shows that six of them (i.e., $First\quad Quartile$, $Mean$, $Median$, $Min$, $Standard\quad Deviation$ and $Third\quad Quartile$) have an obvious effect on the best prediction results, indicated by the higher boxplots. This finding coincides with what we found from the small-scale experiment on the Ant project, suggesting that some of distribution characteristics have greater effects on predicting software defect-proneness.

Interestingly, in Table \ref{CFCPDP}, although the best prediction results of $CPDP-IFS_{pure}$ and $CPDP-IFS_{tca}$ are statistically similar ($p$-value $=0.213>0.05$), $CPDP-IFS_{tca}$ outperforms $CPDP-IFS_{pure}$ in the first eight projects, whereas $CPDP-IFS_{pure}$ performs better in ReLink. Overall, the effect size of $CPDP-IFS_{tca}$ is larger than that of $CPDP-IFS_{pure}$ because of the negative $d$ value. That is, the introduction of transfer learning techniques is in large part valuable for defect prediction. On the other hand, $CPDP-IFS_{pure}$ and $CPDP-IFS_{tca}$ are obviously different with respect to the selection of optimal source projects. The source projects of $CPDP-IFS_{tca}$ tend to have a higher defect ratio (refer to Table \ref{projects}). This finding may be useful to decide which candidate projects are more suitable to be training data for a given target project.

\begin{table}\footnotesize
\newcommand{\tabincell}[2]{\begin{tabular}{@{}#1@{}}#2\end{tabular}}
\centering
\caption{A comparison between CPDP-IFS and CPDP in terms of the Wilcoxon signed-rank test and Cliff's Delta. }\label{CFDP-CPDP}
\begin{tabular}{c|c|c} \hline
  $p-value=0.05$ & \multicolumn{1}{c|}{\tabincell{c}{$CPDP_{pure}$ \\(Cliff's delta $d$)}} & \multicolumn{1}{c}{\tabincell{c}{$CPDP_{tca}$ \\(Cliff's delta $d$)}}  \\ \hline
 $CPDP-IFS_{pure}$ & 0.906 (0.231) & 0.442 (0.000)  \\
 $CPDP-IFS_{tca}$  & 0.441 (0.355) & 0.129 (0.140)  \\  \hline
\end{tabular}
\end{table}

\begin{table}\footnotesize
\newcommand{\tabincell}[2]{\begin{tabular}{@{}#1@{}}#2\end{tabular}}
\centering
\caption{A performance comparison between the CPDP-IFS methods in two settings according to the Wilcoxon signed-rank test ($p-value=0.05$) and Cliff's Delta ($d$). }\label{CFCPDP}
\begin{tabular}{c|c|c|c|c|c} \hline
\multirow{2}*{Target} &\multicolumn{2}{c|}{$CPDP-IFS_{pure}$} &\multicolumn{2}{c|}{$CPDP-IFS_{tca}$} &\multirow{2}*{\tabincell{c}{$Sig.p $\\ ($d$)}} \\
 \cline{2-5}
  & Source & Value & Source & Value &
   \\ \hline
    Ant & Apache & 0.46 & Apache & 0.50 & \multirow{11}*{\tabincell{c}{0.213\\ (-0.223)}}\\
    Xalan   & Equinox & 0.50 & Apache & 0.53 &  \\
    Camel   & Apache & 0.34 & Apache & 0.35 & \\
    Eclipse   & Ant & 0.50 & Xalan & 0.57 &  \\
    Equinox  & Xalan & 0.47 & Apache & 0.61 &  \\
    Lucene  & Ant & 0.42 & Zxing & 0.59 & \\
    Mylyn   & Ant & 0.32 & Xalan & 0.34 & \\
    Pde & Apache & 0.33 & Xalan & 0.39 & \\
    Apache & Eclipse & 0.59 & Xalan & 0.49 & \\
    Safe & Eclipse & 0.65 & Equinox & 0.63 & \\
    Zxing & Eclipse & 0.44 & Equinox & 0.37 & \\ \hline
\end{tabular}
\end{table}

To do this, we further analyzed the impact of \emph{DPR} \cite{He: Simplification} (the ratio of the proportion of defective instances in the training set to the proportion of defective instances in the test set, i.e., $DPR=\frac{\% defects(source)}{\% defects(target)}$) on CPDP-IFS for each target project. Considering the similar trend and limited space, we just take $CPDP-IFS_{tca}$ as an example. The great correlation coefficients, in Table \ref{relationship}, indicate a significantly linear correlation between \emph{DPR} and \emph{f-measure}. The other 10 projects present such a strong positive correlation except the Lucene project, where a very low defect ratio (2.9\%) leads to very high \emph{DPR} values. The strong positive correlations indicate that the prediction performance was improved with an increase in \emph{DPR} value within an appropriate range. In other words, an appropriate \emph{DPR} value is beneficial to CPDP-IFS, whereas too large values are unsuitable for CPDP-IFS. For example, 2.5 is selected as an appropriate threshold for \emph{DPR} in our context.

\begin{table}\footnotesize
\centering
\caption{The correlation coefficients between the performance of $CPDP-IFS_{tca}$ and $DPR$ values. ($\ast\ast$ represents a significance at the level of 0.01, and $\ast$ at the level of 0.05) }\label{relationship}
\begin{tabular}{c|c|c|c} \hline
 project & coefficient & project & coefficient \\ \hline

    Ant & $0.893^{\ast\ast}$ &  Xalan   & $0.886^{\ast\ast}$\\
    Camel   &  $0.906^{\ast\ast}$ & Eclipse   & 0.699\\
    Equinox  & $0.918^{\ast\ast}$ &  Lucene  &$-0.822^{\ast}$  \\
    Mylyn   & 0.651 & Pde & $0.6 88^{\ast}$\\
    Apache & $0.925^{\ast\ast}$ & Safe & $0.817^{\ast}$  \\
    Zxing & $0.825^{\ast}$ & &\\ \hline
\end{tabular}
\end{table}

%\begin{figure*}
%\centering
%\includegraphics[width=6in,height=3.5in]{F-measure.png}
%\caption{The impact of DPR on the $CFDP\_TCA$ performance. The x-axis represents the \emph{f-measure} values}
%\label{F-measure}
%\end{figure*}

\begin{figure*}
\centering
\includegraphics[width=5in,height=1.8in]{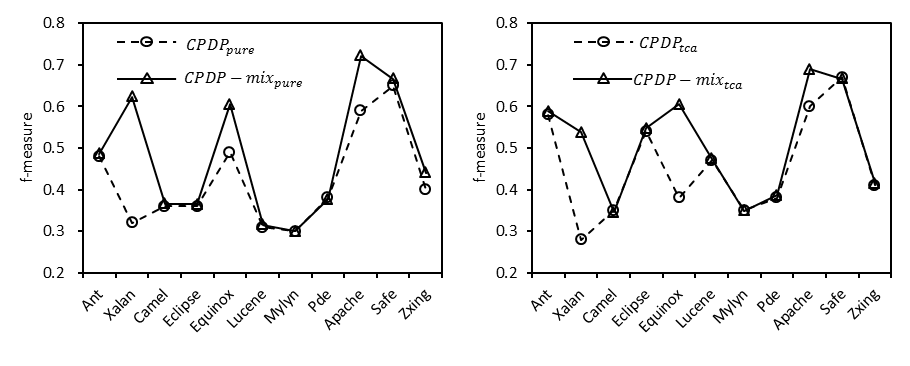}
\caption{The improvement of the hybrid model in \emph{f-measure} in two given settings.}
\label{improvement}
\end{figure*}

\begin{figure*}
\centering
\includegraphics[width=5in,height=1.5in]{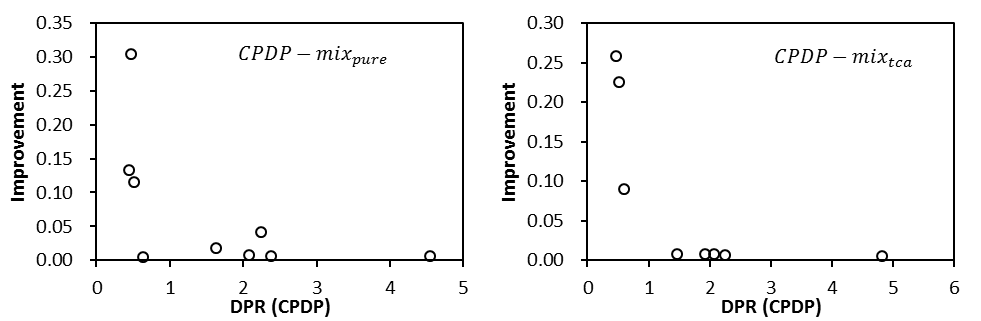}
\caption{The correlations between $DPR$ value and the corresponding improvement in two given settings.}
\label{DPR}
\end{figure*}

Until now, we have validated the practicability and feasibility of CPDP-IFS, which is a rather cheering finding. It suggests that a large number of public defect data sets are no longer limited to specific studies and can be used to the regular CPDP regardless of the difference of metric sets. In the past, one had to engage in tedious metrics-gathering to keep the set of features consistent; in this case, the existing defect data sets can hardly be reused for validating other people's work. For example, due to the different sets of metrics, the authors emphatically declared that the projects in ReLink could not mix with those projects in AEEEM in their studies \cite{Nam:Transfer}. In fact, we not only mixed them but also validated the feasibility of this treatment in this subsection.

\subsection{Can the hybrid model composed of CPDP-IFS and CPDP improve the performance of CPDP?}

Although the above findings have suggested that CPDP-IFS works well and is comparable to CPDP, in general the regular CPDP has not yet passed WPDP \cite{Herbold:Training}. Can the performance of CPDP be improved by the blend combining CPDP with CPDP-IFS? The intuition here arises from the evidence that the defect data from other projects with different metrics may contain more information about software defect from different aspects. For example, source code metrics measure various properties of computer program such as coupling and inheritance, while code churn measures provide an additional perspective on how often code (especially problematic code) is changing over time. So, we can better predict defective instances when these metrics from different aspects are used together.

For this purpose, we built a hybrid model that comprises CPDP and CPDP-IFS to predict the defect proneness of instances. The decision rule of the model is simple, that is, if an instance is classified as a buggy instance (labeled as 1) by either CPDP or CPDP-IFS, it is determined to be defective by the model, whereas the model determines that an instance is defect-free (labeled as 0) only if it is classified as a non-buggy instance by both CPDP and CPDP-IFS. Subsequently, for each target project, we repeated 10 predictions to estimate how well \emph{CPDP-mix} works in terms of \emph{f-measure}. Figure \ref{improvement} shows the best prediction results of the regular CPDP and \emph{CPDP-mix}. There is a significant improvement for one project in each data set, i.e., Xalan (PROMISE), Equinox (ReLink) and Apache (AEEEM), whereas the results of other projects are very stable. In other words, at least the introduction of CPDP-IFS does not have a negative effect on the original CPDP results. Hence, CPDP-IFS can be a sound complement to the regular CPDP.

According to the above finding, we are actually eager to know when the prediction performance of CPDP is more likely to be improved by the hybrid model. Therefore, we analyzed the degree of improvement for each target project in terms of $DPR$. Figure \ref{DPR} shows that the improvement brought by CPDP-IFS is more obvious when the value of \emph{DPR} is very low. More specifically, the value of \emph{DPR} is less than 0.64 in our context. Take the prediction of \emph{CPDP-mix} as an example. The performance improvement of the Xalan project is up to 0.30 when its \emph{DPR} value in the $CPDP_{pure}$ setting is only 0.48. In general, a low \emph{DPR} value indicates imbalanced defect ratios in training and test sets. As we know, the more the defective instances a source project has, the richer the defect information is. This suggests that the hybrid model that introduces CPDP-IFS is helpful to ease the lack of defect information in training data. For example, the $DPR$ values of the Xalan project were increased to 0.86 (close to 1) when using $CPDP-mix_{pure}$.

Despite performance improvements in some specific scenarios, how to select the appropriate distribution indicators to construct a better latent space after non-linear transformation using learning techniques is an interesting future work for CPDP-IFS.

\section{Threats to Validity}
%In this study, we obtained several interesting findings. Meanwhile, potential threats to the validity of our work still remain.

All the three data sets were collected from the Internet. According to the owners' statements, errors inevitably exist in the process of defect identification. For example, there are missing links between bugs and instances in the projects of PROMISE as illustrated in some studies \cite{Bachmann:The,Wu:ReLink}. However, these data sets have been validated and used in several prior studies. Therefore, we believe that our results are credible and suitable for other open-source projects.

We have chosen 11 distinct projects with different sizes and metric sets from three public data sets. However, all projects are written in Java and have been supporting by the communities of Apache and Eclipse. In fact, our experiments should be repeated for more different types of projects. Additionally, we did not introduce any feature selection methods to deal with the 16 indicators when performing our experiments. Meanwhile, we did not utilize any other classifiers except for logistic regression. As a starting point for more general and better CPDP, our approach still has plenty of room for improvement.

The non-parametric statistical test (the Wilcoxon signed-rank test) and Cliff's delta were used throughout our experiments. Other alternative tests can be used when comparing two groups of related samples. In addition, more varieties of effect size measures discussed in literature \cite{Hess: Robust}, such as Cohen's $d$, Hedges' $g$ and Glass' delta, can also be used in our experiments. Even so, we believe that the change of statistical analysis methods does not affect our results. Besides, with respect to the evaluation measure, other commonly-used measures, such as AUC (the area under the ROC curve) and \emph{g-measure} (the harmonic mean of the recall and the specificity), can be used as the criteria to validate the results.

\section{Conclusion}
This study aims to improve the regular CPDP through investigating the problem of imbalanced feature sets between training and test data, and consists of (1) validating the feasibility of our approach to CPDP-IFS proposed in this paper, (2) performing a comparison between CPDP-IFS and the regular CPDP, and (3) testing the ability of the hybrid model combining CPDP with CPDP-IFS to improve the performance of the regular CPDP and providing a guideline for how to select the appropriate source projects when using the hybrid model.

According to the experiments on 11 projects of 3 public defect data sets, the results indicate that our approach based on a distribution characteristic-based instance mapping is comparable to the regular CPDP. Specifically, regardless of the introduction of transfer learning techniques (such as TCA) to CPDP-IFS, our approach can effectively solve the problem of imbalanced metric sets. In addition, the results also show that CPDP-IFS is able to help the regular CPDP when $DPR$ value is every low, and that in some cases, the improvement in \emph{f-measure} is very obvious. In summary, our experimental results show that our approach is viable and practical. We believe that our approach can be useful for software engineers when lower costs are required to build a suitable predictor for their new projects. Moreover, we expect some of our intererting findings could optimize the maintenance activities for software quality assurance.

Our future work will focus primarily on the following aspects: (1) collecting more defect data with different metric sets to validate the generality of our approach; (2) utilizing complex learning techniques to build defect predictors with better prediction performance and capability.

% conference papers do not normally have an appendix

% use section* for acknowledgement
\section*{Acknowledgment}
We greatly appreciate Jaechang Nam and Dr. Pan, the authors of the reference \cite{Nam:Transfer}, for providing us the TCA source program and friendly teaching us how to use it.

This work is supported by the National Basic Research Program of China (No. 2014CB340401), the National Natural Science Foundation of China (Nos. 61273216, 61272111, 61202048 and 61202032), the Science and Technology Innovation Program of Hubei Province (No. 2013AAA020) and the Youth Chenguang Project of Science and Technology of Wuhan City in China (No. 2014070404010232).

% that's all folks
\end{document}